\documentstyle[referee]{l-aa}

\newfont{\mytitlebigfont}{cmr10 scaled \magstep5}
\newfont{\mytitlefont}{cmr10 scaled \magstep3}
\begin{document}
%
%
\def\ang{\thinspace\hbox{\AA}}
\def\km{\thinspace\hbox{km}}
\def\Mpc{\thinspace\hbox{Mpc}}
\def\kpc{\thinspace\hbox{kpc}}
\def\kmsec{\thinspace\hbox{$\hbox{km}\thinspace\hbox{s}^{-1}$}}
\def\kmseg{\thinspace\hbox{$\hbox{km}\thinspace\hbox{s}^{-1}$}}
\def\kmsecmeg{\thinspace\kmsec\Mpc$^{-1}$}
\def\kmsegmeg{\thinspace\kmsec\Mpc$^{-1}$}
\def\ergsec{\thinspace\hbox{$\hbox{erg}\thinspace\hbox{s}^{-1}$}}
\def\ergseg{\thinspace\hbox{$\hbox{erg}\thinspace\hbox{s}^{-1}$}}
\def\ergsqcmsec{\thinspace\hbox{erg}\sqcm\sec$^{-1}$}
\def\sqcm{\thinspace\hbox{$\hbox{cm}^{2}$}}
\def\cm3{\thinspace\hbox{$\hbox{cm}^{3}$}}
\def\cm2{\thinspace\hbox{$\hbox{cm}^{2}$}}
\def\persqcm{\thinspace\hbox{$\hbox{cm}^{-2}$}}
\def\percucm{\thinspace\hbox{$\hbox{cm}^{-3}$}}
\def\kev{\thinspace\hbox{keV}}
\def\sec{\thinspace\hbox{s}}
\def\ha{\hbox{$\hbox{H}_\alpha$}}
\def\hb{\hbox{$\hbox{H}_\beta$}}
\def\hg{\hbox{$\hbox{H}_\gamma$}}
\def\hd{\hbox{$\hbox{H}_\delta$}}
\def\Ha{\hbox{$\hbox{H}_\alpha$}}
\def\Hb{\hbox{$\hbox{H}_\beta$}}
\def\Hg{\hbox{$\hbox{H}_\gamma$}}
\def\Hd{\hbox{$\hbox{H}_\delta$}}
\def\lya{\hbox{$\hbox{Ly}\alpha$}}
\def\Lya{\hbox{$\hbox{Ly}\alpha$}}
\def\kelvin{\thinspace\hbox{K}}
\def\dyncm2{\thinspace\hbox{$\hbox{dyn}\thinspace\hbox{cm}^{-2}$}}
\def\deg{\hbox{$^\circ$}}
\def\rstar{\thinspace\hbox{$\hbox{R}_*$}}
\def\vstar{\thinspace\hbox{$\hbox{V}_*$}}
\def\Zsun{\thinspace\hbox{$\hbox{Z}_{\odot}$}}
\def\msun{\thinspace\hbox{$\hbox{M}_{\odot}$}}
\def\Msun{\thinspace\hbox{$\hbox{M}_{\odot}$}}
\def\rsun{\thinspace\hbox{$\hbox{R}_{\odot}$}}
\def\Rsun{\thinspace\hbox{$\hbox{R}_{\odot}$}}
\def\lsun{\thinspace\hbox{$\hbox{L}_{\odot}$}}
\def\Lsun{\thinspace\hbox{$\hbox{L}_{\odot}$}}
\def\Tsun{\thinspace\hbox{$\hbox{T}_{eff\odot}$}}
\def\zsun{\thinspace\hbox{$\hbox{Z}_{\odot}$}}
\def\gpar{\hbox{$g_{\parallel}$}}
\def\Ca{$\lambda\lambda$\thinspace\hbox{8498,8542,8662}~\AA}
\def\C2{$\lambda\lambda$\thinspace\hbox{8542,8662}~\AA}
\def\Mg{$\lambda$\thinspace\hbox{8807}~\AA}
\def\Fe{\hbox{$[Fe/H]$}}
\def\lg{\hbox{$log\thinspace\hbox{g}$}}
\def\lT{\hbox{$log\thinspace\hbox{T}_{eff}$}}
\def\Teff{\hbox{$T_{\rm eff}$}}
\def\etal   {{\sl et\nobreak\ al.\ }}
\def\integ{\int\limits}

   \thesaurus{       
              (  
               )} 
~

   \title{Calcium Triplet Synthesis }
   \subtitle{ }

   \author{M.L. Garc\'{\i}a-Vargas \inst{1, 2} \and  
Mercedes Moll\'a \inst{3, 4} \and Alessandro Bressan \inst{5} }

   \offprints{M.L. Garc\'{\i}a-Vargas}

   \institute{Villafranca del Castillo Satellite
Tracking Station. PO Box 50727. 28080-Madrid, Spain. \and Present address 
at GTC Project, Instituto de Astrof\'{\i}sica de Canarias. V\'{\i}a L\'actea 
S/N. 38200 LA LAGUNA (Tenerife) \and Dipartamento di
Fisica. Universit\`a di Pisa. Piazza Torricelli 2, I-56100 -Pisa, Italy 
\and Present address at  Departement de Physique
Universit\'e  Laval. Chemin Sainte Foy . Quebec G1K 7P4, Canada 
\and Osservatorio Astronomico di Padova. Vicolo dell' Osservatorio 5.
I-35122 -Padova, Italy.} 

   \date{Received xxxx 1997; accepted xxxx 1998 }

   \maketitle

   \begin{abstract}
\footnote {Tables 1 to 10 are only available in electronic form at CDS via anonymous ftp to cdsarc.u-strasbg.fr (130.79.128.5) or via http://cdsweb.u-strasbg.fr/Abstract.html}

We present theoretical equivalent widths for the sum of the two strongest
lines of the Calcium Triplet, CaT index, in the near-IR ($\lambda\lambda$
8542, 8662 \AA), using evolutionary synthesis techniques and the most
recent models and observational data for this feature in individual stars. 

We compute the CaT index for Single Stellar Populations (instantaneous
burst, standard Salpeter-type IMF) at four different metallicities,
Z=0.004, 0.008, 0.02 (solar) and 0.05, and ranging in age from very young
bursts of star formation (few Myr) to old stellar populations, up to 17
Gyr, representative of galactic globular clusters, elliptical galaxies and
bulges of spirals. The interpretation of the observed equivalent widths of CaT
in different stellar systems is discussed. 

Composite-population models are also computed as a tool to interpret the
CaT detections in star-forming regions, in order to  disentangle between
the component due to Red Supergiant stars, RSG, 
and the underlying, older, population. CaT is found to
be an excellent metallicity-indicator for populations older than
1 Gyr, practically independent of the age. We discuss its application to
remove the age-metallicity degeneracy, characteristic of all studies of
galaxy evolution based on  the usual  integrated indices (both broad band
colors and narrow band indices). The application of the models computed
here to the analysis of a sample of elliptical galaxies will be 
discussed in a forthcoming paper (Gorgas et al. 1997). 


   \end{abstract}

%

\section{Introduction}
The Ca~II triplet in absorption, at $\lambda \lambda$ 8498, 8542, 8662
\AA, is the strongest feature in the near-infrared spectrum of late-type
stars and normal galaxies. Pioneer work by Jones \etal\ (1984) was followed
by those of D\'{\i}az \etal\ (1989, hereafter DTT89), Zhou (1991, hereafter
Z91) and Mallik (1994), who studied the behaviour of these features in
several stellar libraries, as a function of the atmospheric stellar
parameters: effective temperature, T$_{\rm eff}$, surface gravity,
logg, and iron abundance, [Fe/H]. 

DTT89 observed a sample of 106 late-type stars (up to K5III) in the
near-IR, providing the first homogeneous atlas including CaT. They defined
standard spectral windows, free of TiO contamination, to locate the
continuum and to measure the index. They concluded that the equivalent
width of the two main lines ($\lambda \lambda$ 8542, 8662 \AA) of the Ca~II
triplet, EW(CaT), increases with increasing metallicity and decreasing
stellar surface gravity of the star and, at high metallicity (Z $\geq$ 0.5
\Zsun), the surface gravity is the dominant parameter, with values of
EW(CaT) larger than 9~\AA\ found only in RSG. This
behaviour was later confirmed by Z91 and Mallik (1994). 

Zhou reached the same conclusions that DTT89 but adopting slightly
different spectral windows for the index definition. The analysis of the  
values of the EW(CaT)  for the stars in common allow us to combine  
both atlases. In particular, Z91 includes  M-late giant stars,
not present in DTT's library. For these M giants, it appears that the
correlation between EW(CaT) and T$_{\rm eff}$ is stronger than that between
EW(CaT) and log~g. Z91 shows that in M giants EW(CaT) reaches values lower
than the ones predicted only on the basis of the EW-logg calibration
previously found by DTT89. 

Mallik (1994) observed 91 late-type stars, confirming the results of DTT89
and Z91. The lack of the measurement of the 8662 \AA\ line for a large part
of his sample and the different continuum band-passes and spectral
resolution, make very difficult the comparison between Mallik's atlas and
those of  DTT89 and Z91. Mallik's sample does not include stars cooler than
M1, and therefore no conclusions about the values of the index for
extremely cool stars can be achieved. However, for the coolest stars in this
atlas  the values of EW(CaT) are lower for lower T$_{\rm eff}$, confirming
the T$_{\rm eff}$-dependence of the index for the coolest stars. 

Recently, Idiart \etal\ (1997) have published CaT indices for a sample of
55 stars. Their sample do not include cool M stars neither metal-rich
supergiant stars. The cool late-type stars are however included in their calibration since they use those from Z91 (converting the values of CaT given by Z91 to their own system). 
The definition of their indices is different
from the one assumed in this work (common in DTT89 and Z91): they used
different continuum band-passes and the three CaT lines (instead of the
two strongest ones), being the comparison meaningless. Nevertheless, they
confirm the strong dependence on metallicity. With respect to the T$_{eff}$-dependence, they find that the strength of CaT increases from F2 to K5 stars. 

From the theoreical point of view, Smith \& Drake (1987; 1990) and
Erdelyi-Mendes \& Barbuy (1991) computed the intensities of CaT lines, as a
function of the atmospheric parameters (T$_{\rm eff}$, logg and
metallicity). These last authors found that the computed intensity of CaT
lines increases exponentially with metallicity (DTT89 had found a linear
relation but in a narrower range of metallicities), showing a stronger
dependence on metallicity when gravity is low (giant and supergiant stars ). They
also found a weak dependence on effective temperature and a modest 
dependence on gravity. 

Finally, J\o rgensen \etal\ (1992; hereafter JCJ92) computed a complete
grid of models for Ca II lines as a function of  T$_{\rm eff}$,  $logg$  and 
[Ca/H] abundance. They synthesized the equivalent widths of CaT lines. They
used the DTT89 index definition and therefore compared these results with
the published observational data. They found a good agreement between their
calibrations and the observed EW(CaT) compiled by DTT89, reaching basically
the same conclusions already pointed out, that can be summarized as
follows: (1) in high metallicity systems, the stellar surface gravity is
the parameter which controls the strength of the CaT lines; (2) the effect
of the abundance is very important for giants and supergiants, with EW(CaT)
increasing at increasing metallicity, but not for dwarfs; (3) at lower
metallicity the effect of the effective temperature is in competition with
that of the gravity. 

In the present work we compute stellar population synthesis models for the
sum of the equivalent widths of the two strongest lines ($\lambda \lambda $
8542, 8662 \AA ) of the CaT. The age considered ranges from 1 Myr to 13
Gyr, and the metallicity from  0.2 \zsun\ to 2.5 \zsun. Section 2 describes
the main aspects of the evolution related to the appearance of cool stars
on the basis on the Padova evolutionary tracks (2.1), and the computed
Spectral Energy Distributions, SEDs, in which both, stellar (2.2) and
nebular (2.3) contributions have been included. Section 3 is devoted to the
CaT synthesis. Two grids of models have been computed: grid I  which is
based on the theoretical fitting functions of EW(CaT) (section 3.1), and
grid II, based on empirical fitting functions derived from the above stellar
atlases (section 3.2). 

In addition to the models described in section 3, several
composite-population models have been computed with different mass
percentages of young (2.5 -- 5 Myr, able to ionize), intermediate (8 -- 12
Myr, rich in RSG) and very old (10 Gyr) populations. These models are
described in section 4.1 and are meant to constitute a reference frame for
the interpretation of the observations of CaT in star-forming regions at
different scales (from pure HII regions to Starburst galaxies or even
Active Galactic Nuclei, AGN). Section 4.2 discusses the implications of
the use of CaT as a metallicity indicator in elliptical galaxies. Finally, 
section 5 summarizes the conclusions. 

\noindent
\section{Evolutionary Synthesis Models}

We have computed models for Single Stellar Populations (instantaneous
burst), at four different metallicities (Z= 0.2
\zsun, 0.4 \zsun, \zsun\ and 2.5 \zsun, and ranging in age between 1 Myr 
(logt = 6.00) and 13 Gyr (logt = 10.12), with a logarithmic step in
age (given in years) of 0.1. The model at 17 Gyr was also computed 
to compare with other authors (see section 4.2 and Figure 6 ).  

The total mass of the SSP is $1 \times 10^{6}$ \Msun\ with a Salpeter-type
IMF (Salpeter 1955), $\phi(m)=m^{-\alpha}$, $\alpha$ = 2.35, from the lower
limit $m_{low}$ = 0.8 \Msun\ to the upper limit $m_{up}$ = 100 \Msun.
Taking into account the Padova evolutionary tracks (see section 2.1), a
fine grid of isochrones has been computed following the method outlined by
Bertelli \etal\ (1994). We also synthesized a complete grid of isochones
with $m_{low}$ = 0.6 \Msun\ to check the effect of considering a lower
limit of the IMF on CaT index, finding a maximum discrepancy of 10 \%\ and
only for ages older than 4 Gyr. 

Once the HR Diagram is calculated and the SED for the SSP computed (see
sections 2.2, 2.3), we are able to calculate the EW(CaT) in the integrated
populations by taking the EW(CaT) of individual stars from theoretical
models (JCJ92), or observed stellar libraries (DTT89, Z91) as will be
outlined in section 3. 

\noindent
\subsection{Stellar Evolution}

Isochrones were constructed at several ages by interpolating between the
evolutionary sequences calculated by Bressan et al. (1993), and  Fagotto et
al. (1994a,b). These tracks were computed using the radiative opacities of
Iglesias \etal\ (1992) for the initial chemical compositions Z=0.004,
Y=0.24; Z=0.008, Y=0.25; Z=0.02 and Z=0.05, Y=0.352 (Padova models). 

Recent reviews on stellar evolution can be found in Maeder \& Conti (1995)
and Chiosi \etal\ (1992). Here we will briefly summarize the main
properties of the adopted models, with particular emphasis to the red giant
and red supergiant phases which are the most relevant to the CaT synthesis.

Red giant stars appear suddenly after hydrogen in the center has been
exhausted. Two remarkable exceptions are constituted by the most massive
stars if mass-loss is strong enough to peal-off the envelope of the star
thus avoiding the expansion phase, and by  stars around 20 M$_\odot$ if the
mixing criterion in the intermediate convective shell and in the previous
H-burning core is such that the model ignites and burns He in the center as
a yellow supergiant star (case A evolution, usually associated with the
Schwarzschild criterion for the convective instability, Deng \etal\ 1996).
For all the other initial masses the models possess a red giant phase of
significant duration. Old clusters with turn-off mass, M$_{Toff}$, lower 
than 2 M$_\odot$, have a well populated red giant branch (RGB). For a
sufficient high metallicity and/or relatively young age these clusters also
show a red clump of He-burning stars tied to the RGB. On the contrary, in
intermediate-age and young clusters, only the red clump of He-burning stars
is populous and luminous enough to have observable effects. Usually for a
sufficiently high initial mass and low metal content part of the central He
is burnt in a blue loop toward higher effective temperatures. Finally old
and intermediate age clusters, M$_{Toff}$ $<$ 5-6~M$_\odot$, also
display the asymptotic giant branch (AGB) phase. The fuel consumed in this
phase is relatively high so that the contribution to the integrated light
is not negligible. 

As already anticipated the evolution of the most massive stars is still
unclear because of our poor knowledge of the efficiency of internal mixing
processes and of the mass-loss phenomenon. The Padova models account for
mild overshoot from the convective core, and  mass loss by stellar winds
has been accounted for according to the rates given by de Jager \etal\
(1988) from the main sequence up to the so-called de Jager limit in the
HRD. Beyond the de Jager limit the most massive stars enter the region
where Luminous Blue Variables (LBV) are observed and, accordingly, the
mass-loss rate has been increased to $10^{-3}$~M$_\odot$~yr$^{-1}$. As the
evolution proceeds, the surface hydrogen abundance by mass in the most
massive stars eventually falls below the value of 0.3. In this case the
model is supposed to become a Wolf-Rayet (WR) star and the mass-loss rate
is derived according to Langer (1989). 

As a matter of fact there are several unsolved questions in the HR diagram
of the most massive stars, among which we recall the existence of the so
called blue Hertzsprung gap, a region where, contrary to what is observed,
theory predicts a negligible number of stars; the observational evidence of
the de Jager limit at the highest luminosities, which is reproduced by the
models only by adopting an arbitrarily high mass-loss rate (of the order of
10$^{-3}$ M$_\odot$/yr) in the corresponding region of the HR diagram; and
finally the problem of the Wolf-Rayet stars, which are either much cooler
or less luminous than predicted by the models. Nevertheless the theory
predicts that  massive stars with initial mass between
10 and 30 solar masses, spend a significant fraction ($\simeq$50\%) of
their He burning phase as red supergiant but the effective temperature of
these stars is a matter of debate, and this must be reminded when assigning
the spectral type during the synthesis process. In general the effective
temperature predicted by the theory is higher than what is observed, but
one must bear in mind that the majority of the models adopt a static gray
atmosphere as a boundary condition, while that of RSG stars is an extended
and expanding atmosphere. Moreover the suppression of the density inversion
or the adoption of a density scale-height in the convective envelope, both
result into a higher effective temperature (Bressan et al. 1993). Finally
RSG are losing mass at a rate of about 10$^{-5}$ M$_\odot$/yr and dust
processes in the circumstellar envelope can also affect their color and
then their apparent location in the HR diagram. 

Another important question is whether a young SSP may contain RSG and WR
stars at the same time. Bressan (1994) and Garc\'{\i}a-Vargas \etal\
(1995a) have shown that this is marginally possible for an age of 6 Myr and
 Z=0.02. In fact in our standard view, WR stars evolve in the HR diagram
from the highest luminosities almost vertically downward and thus their
presence is associated with very young ages. RSG on the contrary, only
appear after a few Myr have elapsed from the burst onset. However Bressan
(1994) showed that by adopting the mass-loss parameterization of de Jager
et al. (1988),  the predicted mass-loss rate of a typical RSG model of
20~M$_\odot$ of solar composition is  significantly lower than that derived
by means of the Feast formulation (1992), which empirically links the 
mass-loss rate to the period of  pulsation of the RSG stars. Recent models
of massive stars of solar composition, in which one adopts the mass-loss
formulation by Feast (1992), show that 20~M$_\odot$ and 18~M$_\odot$ stars
leave the RSG phase and enter the main sequence band with a surface
hydrogen abundance of 0.43, which is comparable to the one selected by
Maeder for the BSG-WNL transition (Salasnich et al. 1997). This
``horizontal'' evolution into the channel of the low luminosity WR stars allows the presence of WR and RSG stars simultaneously in an instantaneous burst. 

Clusters of intermediate age (between 0.1 and 1 Gyr) are
characterized by the presence of the very luminous Asymptotic Giant Branch
stars (AGB). While their life-time is quite short (around 1 Myr), they are
among the brightest stars in the cluster, their fuel consumption is large
and their contribution to the integrated light is significant. The
appearance of the AGB phase as the SSP evolves is quite sudden at an age of
100 Myr and causes a jump in the colors, in particular when near infrared
pass-bands are considered (see eg. Bressan et al. 1994 ). The same happens
to the EW(CaT) in clusters of about 0.1 Gyr (see Fig. 4). 

At older ages the contribution of the AGB phase declines while that of the
red giant branch (RGB) becomes more and more pronounced. Above 10 Gyr, red
giants mainly belong to the RGB phase and the integrated light from the AGB
phase has become negligible. 

\subsection{The Stellar Energy Distributions}

We have synthesized the emergent spectrum of an evolving star cluster by
calculating the number of stars in each element of the isochrone and assigning 
to it the most adequate stellar atmosphere model, {\it i.e.} the closest one in
effective temperature and surface gravity. The stellar spectrum has then been
scaled to the luminosity of the corresponding theoretical star in the HRD. 

To build our stellar spectral library we assembled the stellar atmospheres
of Clegg \& Middlemass (1987) for stars with \Teff $\geq$ 50000~K and those
of Kurucz (1992) for stars with 5000~K $\leq$ \Teff $<$ 50000~K. The later
models are available at different metallicities. Since the precise shape of
the spectrum of the hottest stars does not have any influence in the CaT
models presented here, we will not discuss the selected atmosphere models
for them (a detailed discussion can be found in Garc\'{\i}a-Vargas 1996 and
references therein). For the coolest stars, we have used a blackbody
distribution since it can model the level of the continuum at 8600 \AA\
better than Kurucz models (of course the SEDs are not used in any case to
synthesize the features, but to locate the continuum level). As an example
, Figure 1 shows observed stellar spectra together with the corresponding Kurucz model and blackbody
(BB) distribution for some representative spectral types. We have checked quantitatively the differences in the continuum level at 8600 \AA\ between 
these three representations (BB, Kurucz, and observed) finding a maximun discrepancy of 15 \%\ for the coolest RSG, and only 1 \%\ for giants. 

\begin{figure*}
\vspace*{10cm}

\caption{Comparison between different near-IR spectra of cool stars and
stellar atmosphere models. Left panel shows a sequence of giants with
effective temperature decreasing from top to bottom. Right panel shows the
Red Supergiant sequence. Data (lines with higher spectral resolution) are
true stars for the labelled spectral type and luminosity class 
(Danks and Dennefeld 1994). The degraded
spectra correspond to a Kurucz's model of a T$_{eff}$ and logg appropriate
for each given star. Finally, the featureless line is the spectral energy
distribution of a blackbody whose T$_{eff}$ has been chosen to be
equivalent to the assigned Kurucz's model. All the spectra are normalized
at 8800 \AA} 
\end{figure*}

\subsection{The Nebular Continuum}

Because our aim is to build models that can be applied to star-forming
regions, we have computed the continuum nebular emission under the
following hypothesis. 

\begin{figure*}
\vspace*{20cm}
\caption{Ratio between the nebular and the total luminosity as a function 
of the burst age. Panels (a), (b), (c), and (d) show the ratio at 
different wavelengths: 4850 \AA, 8600 \AA, 2.17 $\mu m$ and 2.30 $\mu m$ 
respectively. Different line$-$types are used to show the effect at different 
metallicities: Z=0.004 (0.2 \zsun, dash-dotted line), Z=0.008 (0.4 \zsun, 
dashed line), Z=0.02 (\zsun, solid line) and Z=0.05 (2.5 \zsun, dotted line).}
\end{figure*}

The gas is assumed to have an electron temperature, T$_{e}$, which is
metallicity dependent. The values for $T_{e}$= 11000~K (Z= 0.2 \zsun),
9000~K (Z= 0.4 \zsun), 6500~K (Z=\zsun) and 4000~K (Z= 2.5 \zsun) have been
chosen according to the observational determination of T$_{e}$ in
star-forming regions (for the lowest metallicities), and the average value,
in the age-range 1.5 - 5.4 Myr, given by photoionization models (
Garc\'{\i}a-Vargas \etal\ 1995b) for the highest Z values. The assumed
helium abundance by number is 10 \%. The free-free, free-bound emission by
hydrogen and neutral helium, as well as the two photon hydrogen-continuum
have been included. The atomic data were compiled from Aller (1984) and
Ferland (1980) according to the selected value of T$_{e}$. 

Tables 1 and 2 list the integrated luminosity of the SSPs with different
metallicity at some characteristic wavelengths in the UV (2000 \AA),
optical (4850 \AA, representative of the continuum near H$_{\beta}$), and
infrared (at 2.17 $\mu m$, near Br$_{\gamma}$). A complete set of tables,
including the nebular and stellar contributions separately as well as the
total luminosity for our grid of models at wavelengths of 1400, 2000, 4850,
8600 \AA, 2.17 and 2.30 $\mu m$ and the synthetic SEDs are available upon
request. 

Figure 2 shows the ratio between the nebular and the total luminosity as a
function of the age at four selected wavelengths. At \zsun\ the nebular
contribution in the earliest stages of the burst changes between 20 \%\ in
the optical (H$_{\beta}$) to almost 90 \%\ in the infrared (2.3 $\mu m$).
This effect becomes negligible for evolved SSP (older than 5.5 Myr) when
the production of ionizing photons is negligible. 

However, if a very young (ionizing) burst coexist with a slightly older
population (around 10 Myr), RSG rich, such as in the case of some
star-forming regions (Garc\'{\i}a-Vargas \etal\ 1997), the effect of the
nebular continuum competes with that of the older stellar component, and
some stellar infrared features can be partially diluted. This could be the
case of the CO absorption bands at 2.3 $\mu m$, where the contribution of
the nebular to the total luminosity can be as high as 90\%. The same effect
also applies to the near-IR colors. For example, if we assume two
coexisting  populations (one around 2-4 Myr and the other one around 9-12 Myr)
contributing with similar percentage in mass, the resulting V-K color would
be affected  both  by the nebular continuum of the young burst and by the
stellar continuum from RSG present in the intermediate-age burst. Thus
detailed evolutionary synthesis models, using other constraints, would be
required in order to correctly interpret the photometric observations. 

\section{Calcium Triplet Synthesis}

  We calculate the integrated  equivalent widths for the CaT lines by
combining the individual stars in each evolutionary stage, according to the
theoretical isochrone. To this purpose let $I_{j}$ be the intensity in
absorption of the two lines of CaT for each star, $j$, found in the HR
diagram of an SSP: 

\begin{equation}  
        I_{j} = f_{j} EW_{j}
\end{equation}

\noindent
where $f_{j}$ is the corresponding flux at the wavelength of 8600 \AA\ for
the star in the HR diagram. This quantity is obtained by a linear
interpolation between the two central values of the continuum band-passes
as defined by DTT89. The fluxes come from a suitable stellar
atmosphere model and have been scaled to the luminosity of the corresponding
theoretical star in the HR diagram. EW$_{j}$ is the equivalent width of CaT
for a star in the evolutionary stage $j$, that we assume is known. If
$N_{j}$ describes the number of stars in the evolutionary stage $j$ and N
is the total number of points in the HR Diagram, the synthesized equivalent
width of CaT for an SSP at a given epoch is:

\begin{equation}
 EW (CaT, SSP) = \frac{\sum_{j=1}^{N} I_{j}N_{j}}
{\sum_{j=1}^{N} f_{j}N_{j} + f_{neb}}
\end{equation}

\noindent
where $f_{neb}$ is the nebular continuum at 8600 \AA\ corresponding to the SSP. In the following both theoretical (grid I) and empirical (grid II) fitting
functions have been used to obtain the index as a function of the stellar
physical parameters: T$_{eff}$, logg and abundance. The theoretical stellar
grid of EW(CaT) is from JCJ92, while the empirical library is from DTT89
plus the M type stars from Z91's atlas. We consider EW(CaT) to be
zero for stars hotter than 6700~K which is the observational limit of
DTT89's atlas. 

\subsection{Grid I: theoretical fitting functions} 

JCJ92 computed a complete grid of NLTE models for the equivalent widths of
CaT lines from stars with T$_{eff}$ ranging between 4000 and 6600~K,
$logg$ between 0.00 and 4.00, and calcium abundances between 0.1 and 1.6
solar. From their models, the following fitting functions can be used to
calculate the theoretical value of EW(CaT) as a function of \Teff, $logg$,
and calcium abundance, [Ca/H]= -1.0, -0.5, 0.0 and +0.2 (equations (3),
(4), (5) and (6) respectively). 

\begin{equation}
EW_{-1.0} = -5.03 - 0.136\, logg + 0.304\, log^{2}g + 
4.18\,10^{-3}T_{eff} - 4.10\,10^{-7} T_{eff}^{2} - 
3.14\,10^{-4}logg\, T_{eff}
\end{equation}

\begin{equation}
EW_{-0.5} = -10.28 - 1.83\, logg + 0.493\, log^{2}g + 
7.46\,10^{-3}T_{eff} - 7.08\,10^{-7} T_{eff}^{2} - 
2.20\,10^{-4}logg\, T_{eff}
\end{equation}

\begin{equation}
EW_{+0.0} = -14.25 - 5.00\,logg + 0.703\,log^{2}g + 
1.13\,10^{-2}T_{eff} - 1.09\,10^{-6} T_{eff}^{2}
\end{equation}

\begin{equation}
EW_{+0.2} = -16.00 - 5.88\, logg + 0.811\, log^{2}g + 
1.32\,10^{-2}T_{eff} - 1.27\,10^{-6} T_{eff}^{2}
\end{equation}

\noindent
where $EW_{[Ca/H]}$ indicates the value, in \AA, of CaT index 
(sum of the equivalent widths
from the two strongest lines, at 8542, 8662 \AA), computed with the
continuum band-passes located as in DTT89. [Ca/H] means the calcium 
abundance with respect to the solar value. Conversion
between the metallicity Z of the isochrones and the [Ca/H] index of the
fitting functions is made adopting Z=0.02 for [Ca/H]=0 and by linearly
scaling the index for other metallicies. JCJ92 do not compute theoretical
EW(CaT) for metallicities higher than 1.6 solar. We have assumed the use of
equation (6) for our calculations at Z=0.05 (2.5 \zsun), and therefore the
values of EW(CaT) could be understimated. For metallicities lower than
solar a linear interpolation between the values of the indices given by the
above expressions has been done. 

\begin{figure*}
\vspace*{10cm}

\caption{Comparison between data and models of EW(CaT) in stars. Panel a)
shows the EW(CaT) as a function of the gravity. Open circles represent data
from DTT89. Solid lines correspond to JCJ92's fitting functions for the
values of the effective temperature labelled in the figure. The dotted line
is our fit to DTT89's data, which has been used in the models (grid II).
Panel b) shows the EW(CaT) as a function of the effective temperature for
the coolest stars. For stars cooler than 4000~K, we have extrapolated the
expressions given by JCJ92 for stars with T$_{eff}$ between 4000 and
6000~K. Open circles are the data from Z91.} 
\end{figure*}

\subsection{Grid II: empirical fitting functions} 

In the second grid of models we made use of the  observational data
collected by DTT89 complemented by data of M-late type stars from Z91.
DTT89 provide the following relation between EW(CaT), gravity and stellar
abundance as measured by [Fe/H]: 
\begin{equation}
EW(CaT) = 10.21 - 0.95\, logg + 2.18\, [Fe/H]
\label{eq_89}
\end{equation}
This relation has been adopted for the metallicities Z=0.004 and Z=0.008,
assuming [Fe/H]=0 for Z=0.02.

For models with Z=0.02 and Z=0.05, we have fitted the observational
data of the EW(CaT) as a function of the gravity, following equations (8)
and (9). 

\begin{equation}
EW(CaT) = 13.76 - 2.97\, logg  \,~~~;if~~logg < 2
\end{equation}

\begin{equation}
EW(CaT) =  9.51 - 0.78\, logg  \,~~~~;if~~logg \geq\ 2
\end{equation}

The above relations are shown in Figure 3 a) together with the theoretical
calibrations given by JCJ92 for different effective temperatures.

As already anticipated, for M-late stars we adopted the data by Z91, since
these stars were not included in DTT89's library. The data given by Z91 have
been converted to DTT89's system through the following relation, which has
been obtained  by fitting a linear regression to 20 common stars in Z91 and
DTT89:

\begin{equation}
EW(DTT89) = (0.87 \pm 0.07)EW(Z91) + (0.70 \pm 0.58);
\end{equation}

The resulting final expression adopted for M-late stars (T$_{eff}$ $\leq$ 
4000~K and logg $\geq$ 3.00) is:

\begin{equation}
EW(CaT) = (6.06 \pm\ 1.51)\, 10^{-3}T_{eff} - (14.19 \pm\ 5.31);
\end{equation}

Figure 3 b) shows Z91's data for M-type stars and JCJ92' models, as a
function of the effective temperature. Different curves correspond to
models with different gravity as indicated in the plot. 

\vspace*{0.5cm}
With the above fitting functions we computed the synthetic equivalent
widths for the two main lines of CaT at $\lambda \lambda\ 8542, 8662$ \AA\
at the four selected metallicities: 0.2 \zsun, 0.4 \zsun, \zsun\ and 2.5
\zsun. The results are shown in figure 4 and the values are given in tables
3, 4 (grid I), 5 and 6 (grid II). For each table, column (1) lists
the logarithm of the age of the SSP (in yr), column (2) the continuum
luminosity (in units of \Lsun) from the SSP (nebular emission not
included), taking an average value in the DTT89's spectral band-passes;
column (3) the luminosity, in units of \Lsun, absorbed in the Ca II lines
at 8542 and 8662 \AA\ by the stars of the SSP, and column (4) the
equivalent width of CaT, in \AA, computed as the ratio between column (3)
and the total continuum luminosity (in which both the stellar and the
nebular contribution are taken into account). Columns (5), (6) and (7) are
the same of (2), (3) and (4) respectively, but for a different metallicity.

\begin{figure*}
\vspace*{20cm}
\caption{Computed models for the CaT index as a function of the age of the SSP
in a logarithmic scale. Panels a), b), c), and d) display the results for
metallicities 2.5 \zsun, \zsun, 0.4 \zsun, and 0.2 \zsun\ respectively.
Solid points correspond to grid I, and therefore based on JCJ92's theoretical
calibrations for EW(CaT); and open circles correspond to grid II, based on data
from DTT89 and Z91.} 
\end{figure*}

\section{Discussion}

\vspace*{0.5cm}
Figure 4 shows the computed values of the CaT for SSPs. At Z = 2.5 \zsun\
(Fig. 4a) both grids predict similar values of EW(CaT). However, at lower
metallicities, the empirical calibration (grid II) provides EW(CaT) that
are systematically larger than those computed with the theoretical fitting
functions (grid I). For ages older than 100 Myr, the average differences 
between both grids are 1.2, 1.5 and 2 \AA\ for metallicities \zsun, 0.4
\zsun\ and 0.2 \zsun\ respectively. 

JCJ92 suggested that the
differences found between their models and DTT89's data could be due to the
different abundance scale -- we must remind that grid I scales the abundances with [Ca/H] since grid II does with [Fe/H] --  In the present models, a solar abundance ratio
[Ca/Fe] has been assumed, but this could not be the case. In fact, both
observations and  chemical evolution models  show that for low abundances
([Fe/H] $\leq$ -1) the $\alpha$ elements are enhanced with
respect to the solar partition. In particular, the behaviour of [Ca/Fe]
versus [Fe/H] is  shown in Figure 5c from Moll\'{a} \& Ferrini (1995) 
for the galactic bulge: [Ca/Fe] keeps constant 
($\simeq$0.5) for a low  [Fe/H] abundance and thereafter it decreases
towards the solar value. 

The observed enhancement of alpha-elements is due to a lower
proportion of iron group elements to alpha-elements at low Z  when compared
with the corresponding  ratio at the solar Z value. In other words,
stars of low Z (where Z is representative of alpha-elements) have a lower
value of Fe/H than stars with higher Z, as clearly demonstrated by the
study of globular clusters. In particular stars with subsolar Z have
supersolar abundance ratios. To account for this in the comparison
between the observational and predicted values of CaT, we should use
stars of lower observed Z than the value of Z used in the theoretical
isochrones. At lower 
metallicities,  this effect is larger. We also performed
several tests aimed to clarify the role of the [$\alpha$ /Fe] on the
evolution of the star in the HR diagram and they confirmed that isochrones with
the same global metallicity Z but a different enhancement of the
$\alpha$ elements are almost indistinguishable in the HR diagram (see also
Salaris \etal\ 1993). In summary, to compare both grids, we should
use  a non solar  partition of the heavy elements for  abundances lower
than solar. The net effect would be a correction in the values of [Fe/H] adopted in equation (7). 

The real effect of different [Ca/Fe] ratios has been taken
into account by  Idiart etal (1997). These authors, by measuring the CaT
index in a sample of stars whose [Ca/H] and [Fe/H] were known, found a weak dependence of CaT index with the [Ca/Fe] ratio. 

In the galactic star sample used by DTT89 this effect only appears at low metallicity. Therefore it explains 
the differences between DTT89 and JCJ92 results found in panels c) and d), because the low abundance stars present in DTT89´s sample have been used to compute our grid II. 
However, the same explanation cannot be invoke in the case of solar abundance, panel b), where the partition must be solar for the neighbourhood stars. 

The differences found between the two grids in the 
oldest populations at Z = 0.008 could be due to the use, in grid II, of the
solar M-late relation, equation (11), also at Z=0.008, producing values
of EW(CaT) that could be overestimated. This does not occur in grid I, in
which both  T$_{\rm eff}$ and abundance dependence are consistently
taken into account in the theoretical calibrations. For these reasons we
consider  grid I more reliable than grid II although, on the other hand, this 
last one rest on the extrapolation of the JCJ's relation for the coolest stars, 
for which unfortunately, we have not found observed values either theoretical models.

The above disagreement between the two grids notwithstanding, we may draw
the following general conclusions.

At the higher metallicities the EW(CaT) shows a clear maximum around 10
Myr. This is due to the prominence of the RSG phase at these ages and metal
content. At earlier stages massive stars evolve according to the O-BSG-WR
sequence, while at lower metallicity the scheme followed is  O-BSG-YSG-RSG, 
with the later phase being only a tiny fraction of the total lifetime. 

The dependence on metallicity can be easily quantified. Synthetic values of
the index higher than 7 \AA\ are only found in models with metallicity
\zsun\ or higher, reaching values as high as 11 \AA\ only for 2.5 \zsun\
models. The variation with the  metallicity is due, on one hand, to the
intrinsic dependence of the index  and, on the other, to the stellar
evolution effect just described (see also section 2.1) 

From this maximum value the index decreases as the age increases up to a 
value of about 100 Myr. In the case of the two metal poor sets the index
remains almost constant with time up to this age. Around 100 Myr the
appearance of the AGB phase produces a sudden increase of the index which
then decreases until 1 Gyr. At this stage the advent of the RGB
induces another discontinuity which is more evident at the highest
metallicity. As the increasing duration of the RGB phase at increasing age
is compensated by a decrease of the evolutionary flux of stars and by a
shortening of the AGB phase, the integrated value of the index becomes
almost age-independent. In clusters older than a few Gyr the metallicity is
the dominant parameter driving the integrated value of EW(CaT). 

\begin{figure*}
\vspace*{10cm}
\caption{EW(CaT) versus [Fe/H]. Filled symbols are data measured in old
galactic globular clusters compiled from the literature (Bica \& Alloin
1986b, 1987; Armandroff \& Da Costa 1991; Armandroff \etal\ 1992 and
Geisler \etal\ 1995). Open symbols are our two grids of models as labelled
in the plot.} 
\end{figure*}

It would be desirable to compare these models with the equivalent widths of
clusters at different ages and metallicities  (SSP) making use of the same
isochrones library. 

The main body of available data is the one from Bica \& Alloin (1986a, b)
who presented a data-base of star clusters at different ages and
metallicities. For the young metal-poor clusters in the Magellanic Clouds,
Bica \etal\ (1986, 1990) give also the value of the CaT, but the error
bars quoted for the age and metallicity are too large to provide a
reliable test for our models. 

In the case of  old SSPs, for which we have shown  that the CaT index is
mainly a function of abundance, we have collected in Figure 5 the observed
values of the EW(CaT) against [Fe/H] for several globular clusters and
we have compared them with the results from our models at an age of 13~Gyr. We must refer our results to [Fe/H] abundance scale. This is the case of grid II. However, since grid I uses [Ca/H], we must account for the enhacement of [$\alpha$-elements/Fe] as we have explained above. Therefore, we have assigned the value of [Fe/H] for every total abundance Z or [Ca/H] value, 
by using the [Ca/Fe] relation found by Moll\'a \& Ferrini (1995) already quoted. This relation implies a correction of 0.0, -0.2 and -0.4 dex for values of [Ca/H] solar, 0.4 solar and 0.2 solar respectively. 

Data in Figure 5 shows that in old systems a narrow correlation
between the CaT index and the metallicity over more than two orders of
magnitude in [Fe/H] is found and, at the same time, they provide a significant
reliability test for the theoretical models presented here. 

Finally, a recent paper by Mayya (1997) presents CaT synthesis models to be 
applied to starburst regions. Therefore only young population results
may be compared with our models. Mayya uses JCJ92 fitting functions for Z$\leq \zsun$ and DTT89 empirical relations for higher metallicities. He uses the stellar evolutionary tracks from Geneva group. His results also show a primary peak due to the RSG phase, a secondary maximum and a low constant value for 
SSP older than 100 Myr. Both peaks occur at earlier ages than in our
models, due to differences in the assumed stellar tracks, and the
asymptotic value is lower than the one in our grid II. The
most important difference appears at lower abundances: at Z=0.008 the
first maximum disappears in our models,  while it exists in Mayya's. The evolutionary tracks selected by
Mayya (1997) with enhaced mass loss rates for low 
abundances produce this behaviour, not predicted with the Padova evolutionary models either previous generation of Geneve tracks. The convenience of the use of these enhanced mass loss rates is still a matter of debate.

\subsection{CaT synthesis in composite-populations: unveiling the 
presence of RSG in star-forming regions}

CaT has been observed not only in star clusters and normal galaxies but
also in Active Galaxies (Terlevich \etal\ 1990a; Nelson \& Whittle 1995;
Palacios \etal\ 1997) and star-forming regions like Starbursts
(Terlevich \etal\ 1990a, b; Garc\'{\i}a-Vargas \etal\ 1993; 
Gonz\'alez-Delgado \etal\ 1995) and 
Giant Extragalactic HII Regions, GEHRs, (Pastoriza \etal\ 1993; 
Gonz\'alez-Delgado \etal\ 1995; Terlevich \etal\ 1996). 

There is a controversy related to the origin of the observed CaT in
star-forming regions and AGNmostly because these regions are not
spatially resolved from the ground. Therefore one of the key
questions is if the observed CaT comes from a single stellar population RSG rich or from  a result of a mixture of populations of varying age and possibly metallicity (including the RSG plus the underlying older population). 

In the case of isolated GEHRs, and therefore not contaminated by an
underlying old population, two possibilities can arise: (1) the
production of the CaT is due to the same young burst that is ionizing the region and (2) the CaT is produced in a slightly older (10-15 Myr) population, coexisting in the same GEHR with the younger, ionizing, one. With respect to the first scenario, current theoretical models (Salasnich \etal\ 1997) predict a narrow range of age and metallicity in which an SSP can produce both ionizing stars (O and WR stars) and RSG, namely around 4-6 Myr and at solar metallicity . This has been proposed for the CaT observations in NGC 604 (Terlevich \etal\ 1996). However, some other GEHR need the existence of an older component, second scenario, such is the case of some GEHR in the circumnuclear region of NGC7714 (Garc\'{\i}a-Vargas \etal\ 1997). 

The largest circumnuclear GEHRs usually show the CaT feature in their
spectra. However, some contamination from the older underlying population
in the host galaxy is expected and therefore it is not clear whether  the
CaT is originated in the GEHR or in the disk-bulge population
(Garc\'{\i}a-Vargas \etal\ 1997). 

 In the case of starburst galaxies and AGN the picture is even more
difficult to interpret, and the need for models which include the CaT
synthesis from different populations becomes a key issue. To study  this 
problem we have computed composite models with a combination of
three different kind of populations: (a) a young one, able to ionize the
gas, and definitively present in the region, (b) an intermediate age one,  
RSG rich, and (c) a very old population representative of those 
present in ellipticals and bulges of spirals. The selected ages are, 2.5
and 5 Myr for the youngest population, 8 and 12 Myr for the intermediate
component and 10 Gyr for the oldest one. Three types of models have
been computed: (1) a combination of two coexisting bursts, young, and
intermediate, contributing 50~\% each in mass, suitable to be used in 
GEHRs, without any underlying population; (2) a two-component
model in which the young burst plus the old population are combined in different
proportions, and (3) a three-component model in which two coexisting bursts,
young and intermediate-age, plus the old underlying population are contributing 
to the light in different percentages. The metallicity of the old population has been
chosen to be \zsun, 0.4 \zsun\ and 0.2 \zsun\ for young populations with
2.5 \zsun, \zsun\ and 0.4 \zsun\ respectively, according to what is predicted by 
chemical evolution models. 

To define the relative proportions  we use the ratio, P, of the continuum
luminosity at 6500 \AA\ (close to \ha) of the young and intermediate
population (when present) to the total light. As an example P = 
0.10 indicates a model in which the population characteristic of  the region
(young or young + intermediate) is contributing 10 \%\ to the total light
in the continuum at \ha\ . This method allows a check of the adopted
proportions by a direct inspection of the \ha\ images. We have computed models
with P ranging from the ones typical of GEHRs (P = 0.10 - 1  going from the smallest
to the largest regions) to the ones characteristic of the most powerful
starburst galaxies (P= 1 - 100).

Tables 7, 8 (grid I) 9 and 10 (grid II) display the results of
the composite-population models. Each table contains the results for
three metallicities: 0.4 \zsun, \zsun\ and 2.5 \zsun. The first column lists
the proportion, P, defined above (including the two -- tables 7, 9 --- or
three --- tables 8, 10 --- populations considered). If  P is not given, a
single population, or a combination of two coexisting young populations
contributing 50 \% in mass each, have been considered. Column 2 shows the age
of the population(s), in Myr. Therefore 2.5 + 10$^{4}$ correspond to a model in
which a young burst of 2.5 Myr is combined with an old population of 10 Gyr.
Column 3, EW(CaT), lists the value of the equivalent width of CaT in
absorption, in \AA, for each model. Finally column 4 is the equivalent width of
\hb\ Balmer line in emission. If this value is missing then the adopted
population(s) is(are) too old to produce ionizing photons.

The predicted values of EW(\hb) in emission have been computed without
considering the dust associated to the ionized region. There exists a well
known discrepancy between predicted and observed values of EW(H$_{\beta}$)
(e.g. Viallefond \& Goss 1986). In fact, only 3 out of 425 HII galaxies in
the catalogue by Terlevich \etal \ (1991) show EW(H$_{\beta}$) comparable
to the ones calculated for clusters younger than about 3 Myr (i.e. $>$ 350
\AA; Mas-Hesse \& Kunth 1991; Garc\'{\i}a-Vargas \etal\ 1995a; 
Stasi\'nska \& Leitherer 1996). Under the assumption of a single burst
population and a radiation bound nebula, an explanation for this
disagreement could be that the reddening affecting the emission lines is
caused by dust inside the regions (associated to the gas) which  therefore
does not affect the continuum of the ionizing cluster (Mayya \& Prabhu
1996). If this is the case, the measured EW(H$_{\beta}$) should be
increased according to the reddening determined from the emission lines and
taking into account the contribution of the nebular continuum (
Garc\'{\i}a-Vargas \etal\ 1997). 

Columns 5, 6 ( \zsun) and 7, 8 (0.4 \zsun) contain 
the same as columns  3, 4  already described for the case of 2.5 \zsun. 

\begin{figure*}
\vspace*{20cm}
\caption{Diagnostic Diagram of EW(CaT) -- in absorption -- versus
EW(H$_{\beta}$) -- in emission -- used as a tool to unveil the presence of
RSG in star-forming regions. Triangles correspond to three component models, 
and asterisks are the two-component models, in which the CaT is contributed by the old population (10 Gyr in the models). 
Additionally, squares indicate models in which only a young and an intermediate component are considered. These two latest models should be applied to isolated GEHRs in which if 
CaT was detected it would be necessarily due to the presence of RSG.}
\end{figure*}

To summarize the results, we plot in Figure 6 the value of EW(CaT) against
that of  EW(H$_{\beta}$). This figure can be used as a diagnostic diagram
to unveil the presence of an intermediate population RSG rich,
when analyzing data of star-forming regions which are located over
an older underlying population. This is usually the case of  AGN, nuclear
starbursts and circumnuclear GEHRs. The results  for the grid I are divided in in four panels. 
These panels simulate four different types of star-forming regions: a)AGN, corresponding to solar metallicity and large P values, b) circumnuclear high metallicity HII regions with solar metallicity but low P values, c) nuclear starbursts having half solar metallicity  and high
P, and d) circumnuclear GEHR with moderate metallicity (0.4 \zsun) and low
P values. 

Inspection of tables 7, 8 and Figure 6 shows that for powerful starburst
galaxies (with values of P larger than 1.00) values of EW(CaT) higher than
3.8 \AA\ are predicted only if RSG are present in the region and, in this
case, the two-component models (ionizing burst+bulge) would not be able to
reproduce the observations. Garc\'{\i}a-Vargas \etal\ (1993) gave values of
CaT in starbursts ranging between 2.5 and 8 \AA. As can be seen in Figure 6
c) a value as low as 2.5 \AA\ can imply a low metallicity, an older age or
simply the absence of RSG, and only a detailed study of other
observational constraints could provide the solution. On the contrary, values
as high as 8 \AA\ would necessarily imply the presence of RSG and a
metallicity at least solar, therefore somewhat higher than the average
metallicity found in starburst galaxies. 

The same method can be applied to AGN. In this case, a detection of CaT
higher than 5 \AA\ implies the presence of RSG inside the region sampled by
the slit , probably larger than the nucleus and including also the subarcsec
circumnuclear rings as shown by HST observations of  some of the nearest
AGN (Colina \etal\ 1997). Terlevich \etal\ (1990a) showed that all AGN in 
their sample had values of EW(CaT)~$\geq$~5 \AA, therefore implying the
presence of RSG. 

In the case of circumnuclear GEHRs the discrimination between the presence
or absence of RSG is a difficult task, particularly at moderate
metallicity (see Fig6b and 6d). A more detailed study with further
observational constraints is needed to discriminate between the two
possibilities, such as the analysis of the whole optical spectrum to
constrain the age of the young burst and  an image near \ha\ to determine P
(Garc\'{\i}a-Vargas \etal\ 1997) 

\subsection{CaT in old populations: a strong metallicity constraint}
 
We now turn to old populations, namely elliptical galaxies and bulges of
spirals. Terlevich \etal\ (1990a) present a sample of 14 objects, whose
EW(CaT) are between 6.1 and 8.1 \AA\ (typical error bar of $\pm$ 0.8 \AA)
measured as in DTT89 and thus directly comparable to our models. 
Delisle \& Hardy (1992) give central values (and also
gradients) for 12 galaxies, with CaT equivalent widths
ranging between 6.4 and 7.7 \AA\ (typical error bar of $\pm$ 0.2 \AA). In
spite of  different spectral band-passes than DTT89,  but also free from 
TiO bands contamination,  the comparison of  three common objects, M~31,
M~81 and NGC~1700, gives values of 6.4, 7.3 and 7.0 \AA\ in Delisle \&
Hardy (1992) and 6.1, 7.7 and 6.1 in Terlevich \etal\ (1990a) respectively,
which are consistent within the errors. In summary the available observed
values of the CaT index in old populations (elliptical and bulges of spirals)
are between 6 and 8 \AA.  These numbers compare well with our old SSP 
models of solar metallicity (Fig. 4b) and suggest a quite uniform average 
metallicity for these systems in agreement with what is derived by detailed 
galactic models of narrow band indices (Bressan et al. 1996). 

Vazdekis \etal\  (1996, V96) compute evolutionary synthesis
models for early-type galaxies, with metallicities 0.4 \zsun, \zsun, and
2.5 \zsun\ and ages 1, 4, 8, 12 and 17 Gyr. They consider different hypotheses
 about the IMF, the
chemical evolution and the star formation history, producing a set of
models which includes colours and line indices, in particular CaT.  
Since the evolutionary scheme is the same that the one 
assumed in our models, we present in Table 11 a comparison between V96 
models and our grid II for the SSP with common ages and metallicities. 

\setcounter{table}{10}
\begin{table}
\caption[]{Comparison between V96 and grid II for old populations}
\begin{flushleft}
\begin{tabular}{ccccccc}
\hline
\noalign{\smallskip}
Z & Model & 1 Gyr & 4 Gyr & 8 Gyr & 12 Gyr & 17 Gyr \\
 \noalign{\smallskip}
\hline
\noalign{\smallskip}
 0.008 & V96     & 6.24 & 6.63 & 6.79 & 6.88 & 6.97  \\
 0.008  & grid II   & 5.78 & 7.21 & 7.31 & 7.43 & 7.48  \\
 \hline
 0.02   & V96      & 8.32 & 8.44 & 8.25 & 8.22 & 8.09  \\
  0.02   & grid II   & 6.84 & 8.57 & 8.57 & 8.63 & 8.57  \\
\hline
  0.05 & V96        & 8.50 & 8.90 & 8.34 & 8.08 & 7.92  \\
  0.05  & grid II    & 7.22 & 9.92 & 9.62 & 9.65 & 9.63  \\
\noalign{\smallskip} 
\hline
\noalign{\smallskip}
\end{tabular}
\end{flushleft}
\end{table}

V96 give values higher than grid II at Z=0.02 and 0.05, and lower at Z=0.008 
(except at 1 Gyr) . For the highest metallicty, 
a source of the discrepancy could be the different assumed fitting function 
for the index (although both based on DTT89, they use a single fit for any 
abundance, as in DTT89,  since we use equations 8 and 9 for \zsun\ and 
2.5  \zsun\ . The rest should be due to the inclusion in our models  
of the coolest late-type stars, Z91, and therefore a different modelization of 
CaT index for cool stars, important in old populations. This comparison stress the 
need of observations of cool stars to test the present calibratios.
 
Idiart etal (1997)  also compute synthetic values of CaT in old populations. They use a calibration based on their own star sample, with a different index definition, which includes the three calcium lines. They also include in their models late-type stars from Z91 but not high metallicity RSG stars (although these stars are not present in such old populations, the lack of high metallicity RSG in their star sample implies that their values would be definitively lower than ours in RSG-rich populations at metallicities solar or higher than solar). In the range-age (1-12 Gyr) in which we can compare our models with Idiart's, their resulting EW(CaT) for SSP are slightly lower than those from our grid II models. However, the values agree quite well (as an example, at 0.3 \zsun\ their values are 5.53 and 6.50 \AA\ for populations at 1 Gyr and 12 Gyr, since ours (as calculated as an average value of 0.2 and 0.4 \zsun) are 5.69 and 7.19 \AA\ respectively. At \zsun\  Idiart's values are 6.57 and 7.55 \AA\ again for the extreme ages 1 Gyr and 12 Gyr, since ours are 6.84 and 8.60 \AA\ at the same ages. Differences can come not only from the calculations in the CaT index but from the different assumptions adopted in the low-mass limit of the IMF and in the evolution of the low-mass stars. 

An important point is the relatively low sensitivity of the
CaT index to the age above a few Gyr, makes it a very powerful tool for
discriminating among the metallicity of the stellar systems. 
It is well known that due to the similar response of the isochrone turn-off
to variations in age or metallicity it is difficult to disentangle age and
metallicity effects by the sole analysis of the integrated properties of
the spectra in old populations. Different diagnostic diagrams have been
adopted, as can be seen in  Gonz\'alez (1993) and Bressan \etal\ (1996)
none of which is fully adequate to overcome this difficulty. 
The quantity 
\begin{equation}
	\frac {\delta log CaT/\delta (log age) } {\delta log CaT/\delta log Z} 
\end{equation}
is a measure of the relative sensitivity to age and metallicity. At
t=13~Gyr and Z=\zsun\ this quantity is 2.876, but the average value between
2 and 13 Gyr is 6.4. Among the narrow band indices considered so far in the
literature (Gorgas et al. 1993, Worthey et al. 1994, Bressan et al. 1996)
this is the one with the highest sensitivity to the metal content. Jones \&
Worthey (1995) claimed that the Fe$_{4668}$ index has a large sensitivity
to the metallicity with a value $\delta$ (log age)/$\delta$ (log Z)=4.9, that is lower than
our mean result of 6.4.  These same authors use the index
H$_{\gamma_{HR}}$ as an age discriminator, due to its total
independence of the metallicity: $\delta$ (log age)/$\delta$ (log Z)=0.0. 
We have not calculated it, but a suitable
combination of the CaT index with another whose age sensitivity is higher,
such as the H$_\beta$ index (the following one with low sensitivity to the
metallicity with a $\delta$ (log age)/$\delta$ (log Z)=0.6, could definitively separate age
from metallicity and solve the age-metallicity dilemma in early-type
galaxies (Bressan et al. 1996). 
Thus, as a preliminary step and awaiting for our own complete galaxy models,
we can generate mixed diagnostic diagrams using our models for CaT index and the 
narrow band indices from Bressan \etal\ (1996), that are computed
adopting the same library of stellar evolutionary tracks. 

\begin{figure*}
\vspace*{10cm}
\caption{ EW(CaT) (grid I, this work) versus the computed values for the logarithm of
\hb\ index (Bressan \etal\  1996) in old populations.}
\end{figure*}

Figure 7 shows the synthetic values of the EW(CaT) plotted
against the logarithm of \hb\ index as modelled by Bressan \etal\ (1996). It can be
seen that curves of constant age are almost orthogonal to curves of
constant metallicity, making this diagram one of the most powerful tools to
solve the problem of the age-metallicity degeneracy in elliptical galaxies
in a similar way to Figure 2 from Jones \& Worthey (1995). 

Although more
theoretical work has to be made to assess the importance of the study of 
CaT in old systems, the use of the later index as a straightforward
metallicity indicator in early-type galaxies is very promising (Gorgas
\etal\ 1997). 

\section{Conclusions}

Models for CaT index have been computed for SSP ranging in age from 1 Myr to 17 Gyr and 
in metallicity from  0.2 to 2.5 \zsun\  in order to evaluate the usefulness of this index to 
constrain the age and metallicity of stellar populations. Two grids of models have been generated, 
based on different calibrations of the index as a function of the stellar parameters: effective temperature, gravity and abundance: grid I assumes the theoretical calibration given by JCJ92 and grid II a fitting to  observational data (DTT89, Z91). Values from grid I are systematically lower than those from grid II. Since the abundance scale is different in the two set of models,  [Ca/H] in grid I, [Fe/H] in grid II, the differences at low metallicity can be explained if an enhacement of  $\alpha$ elements, like Ca, with respect to the solar partition is assumed. When cool M-type stars are present in a SSP, the different calibration assumed in both grids leads to different values for CaT, and more data and models 
are needed to account properly for the contribution of these stars to the index.

The evolution of  EW(CaT) in SSP with time presents a primary maximum when RSG appear (around 10 Myr). The value of this peak is strong metallicity-dependent due to both, the evolutionary tracks and the intrinsic dependence of the index with the abundance for supergiants. At about 100 Myr the appearance of the AGB phase produces a sudden increase of the index (secondary maximum) followed by a decrease until 1 Gyr, remaining then almost constant for a given metallicity.

In star-forming regions,  we propose the use of  a diagnostic diagram, EW(CaT) versus EW(\hb) in emission, to evaluate if CaT is produced by RSG in the region, or by an underlying old population and therefore as a tool to date the starburst. This diagnostic is a powerful indicator of the presence of RSG in  high metallicity scenarios, where values larger than 4-5 \AA\ unveil the presence of young RSG. At lower metallicities, the sole use of the diagram cannot disentangle between giant and RSG contributions, especially if the starburst is not very luminous with respect to the underlying population.  

Finally, the behaviour of the index in populations older than 1 Gyr shows that the strength of CaT index is 
controlled by the abundance, leading us to propose another diagnostic diagram, this time using EW(CaT)  versus the logarithm of the traditionally used \hb\  (absorption) index, as a tool to break the 
age-metallicity degeneracy in elliptical galaxies.  

More theoretical work and observed data are needed to assess the importance of CaT for abundance 
and age determination in stellar populations.
 
\begin{acknowledgements}
We thank Javier Gorgas, Enrique P\'erez , Rosa Gonz\'alez-Delgado, Eduardo Hardy and Claus Leitherer for their helpful comments. This work has been partially supported by the
 Spanish DGICYT project PB 93-052 and by the TMR grant ERBFMRX-CT96-0086 
from the European Community. We also thank the anonymous referee for the useful comments which have contributed significatively to the improvement of the paper. Finally, we thank N. Smith for her help with English during the revision of the manuscript.
\end{acknowledgements}

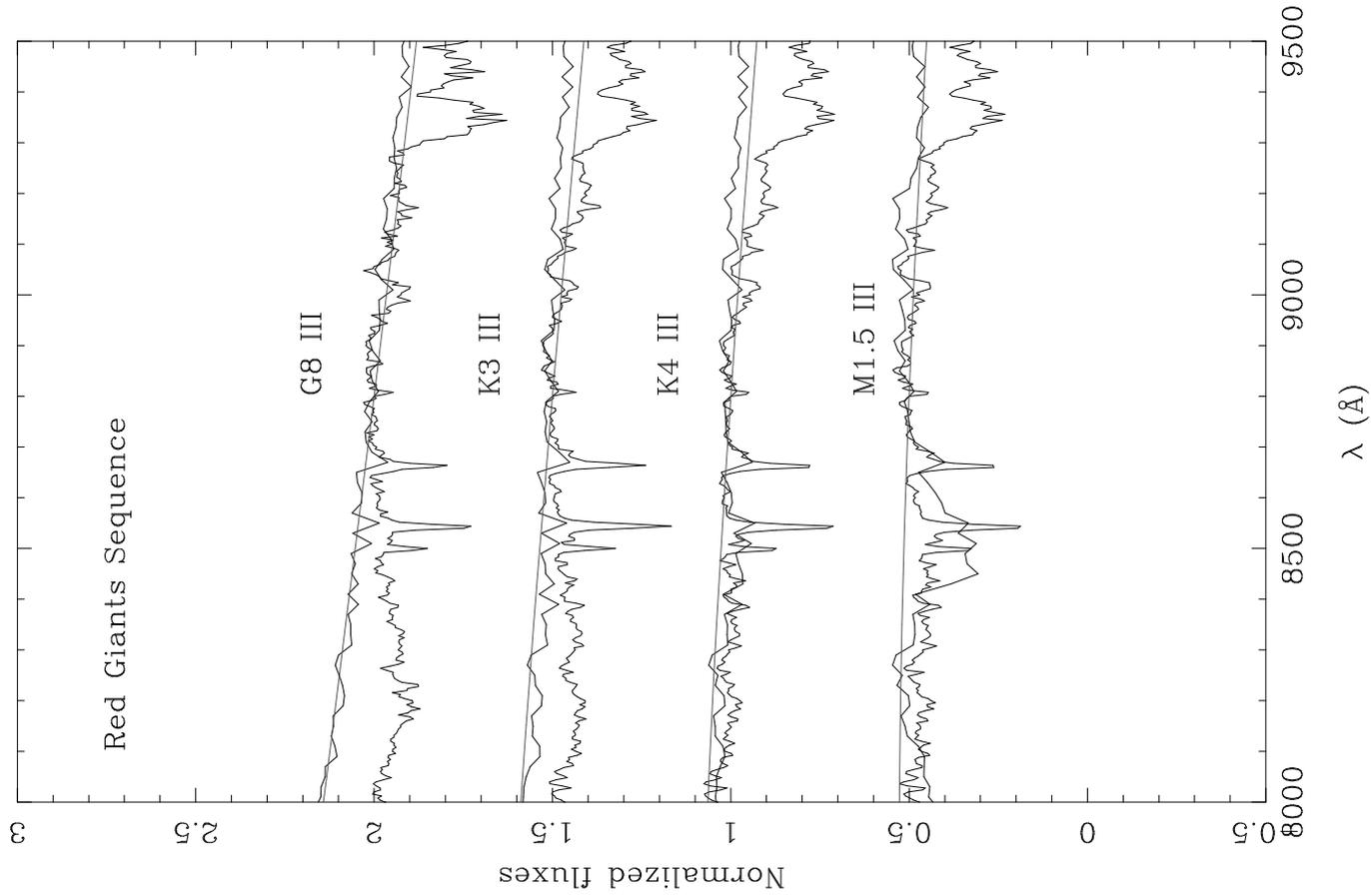
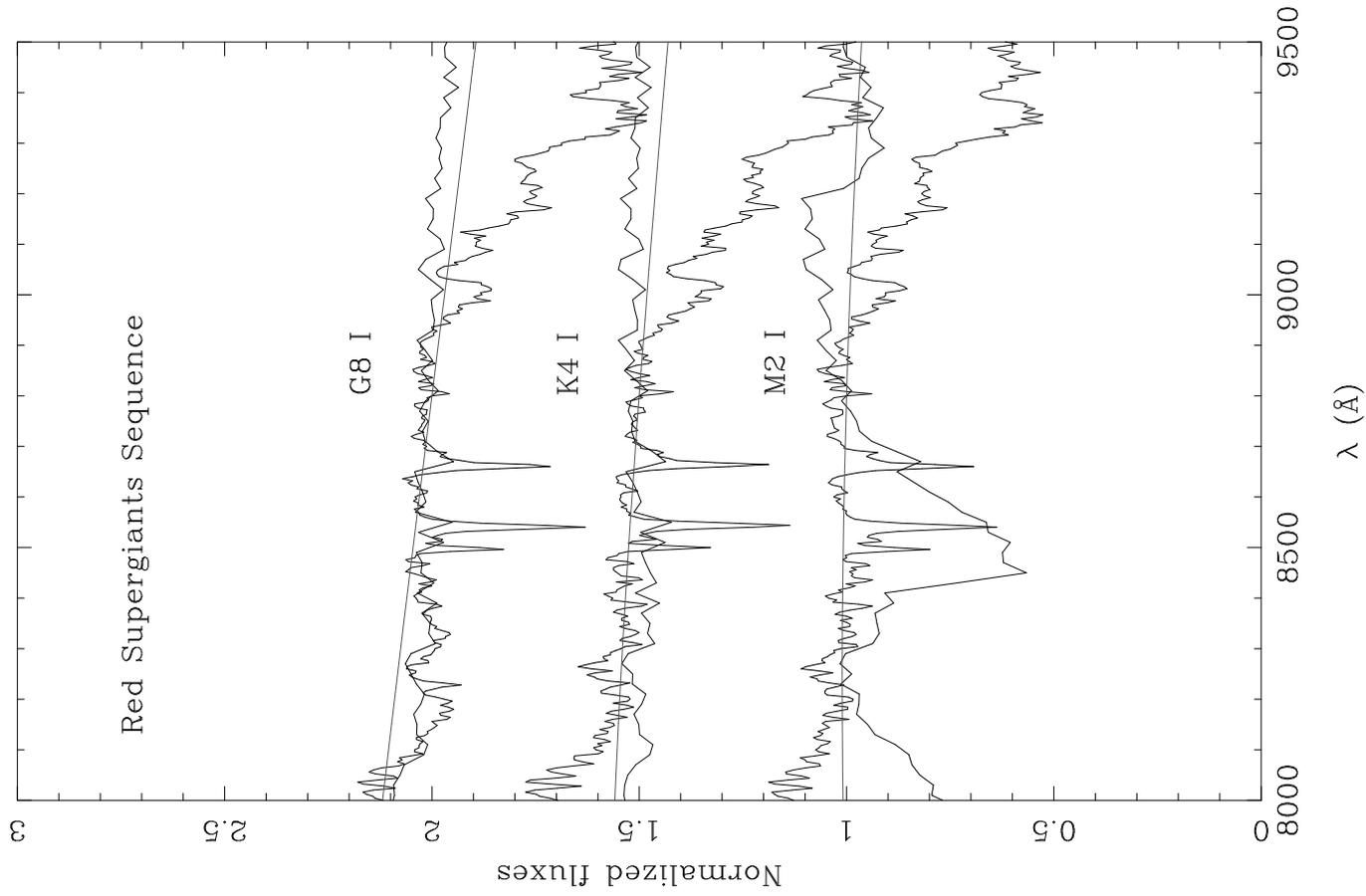
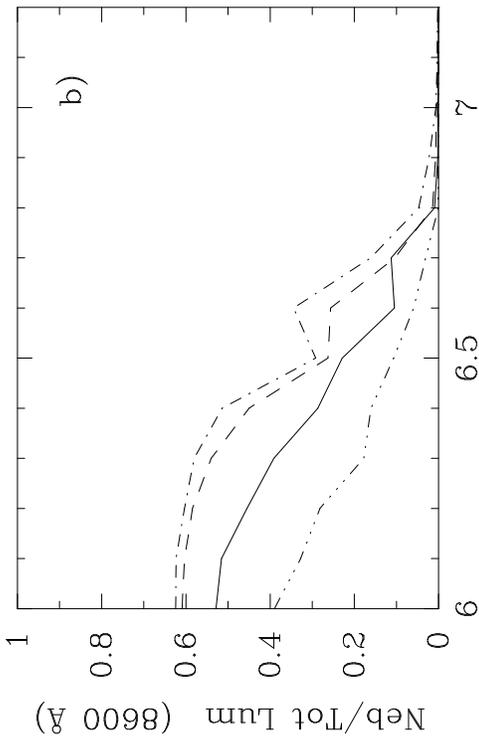
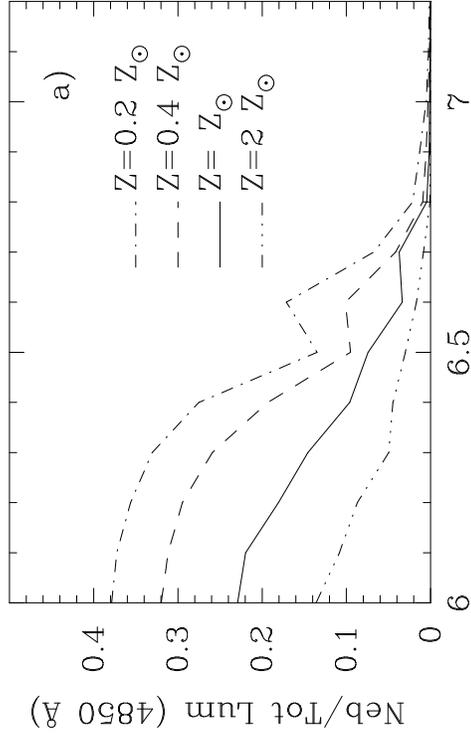
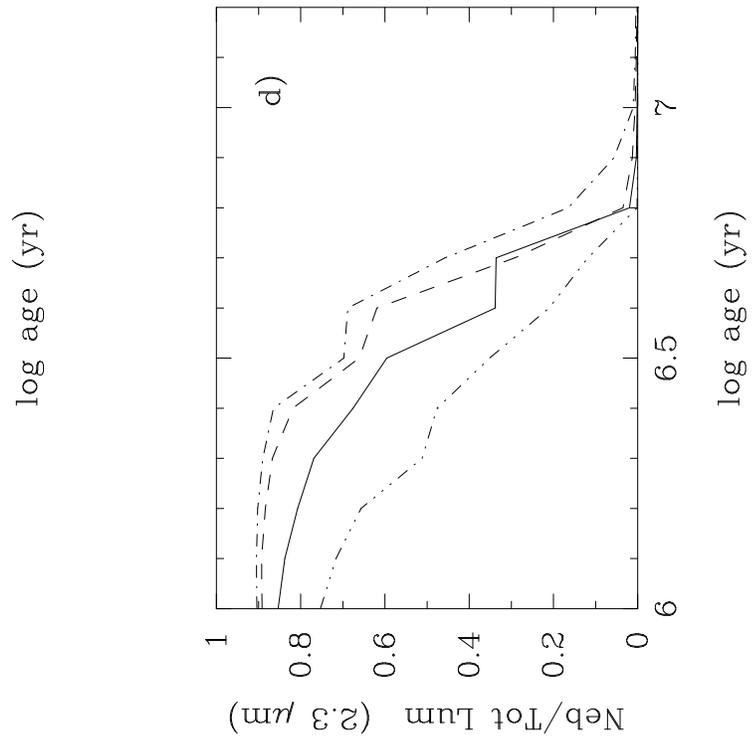
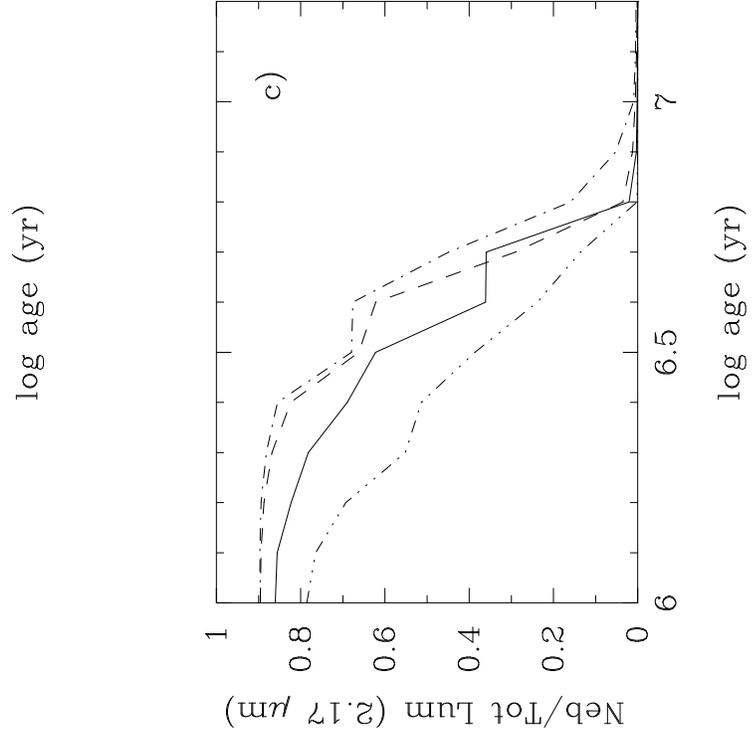
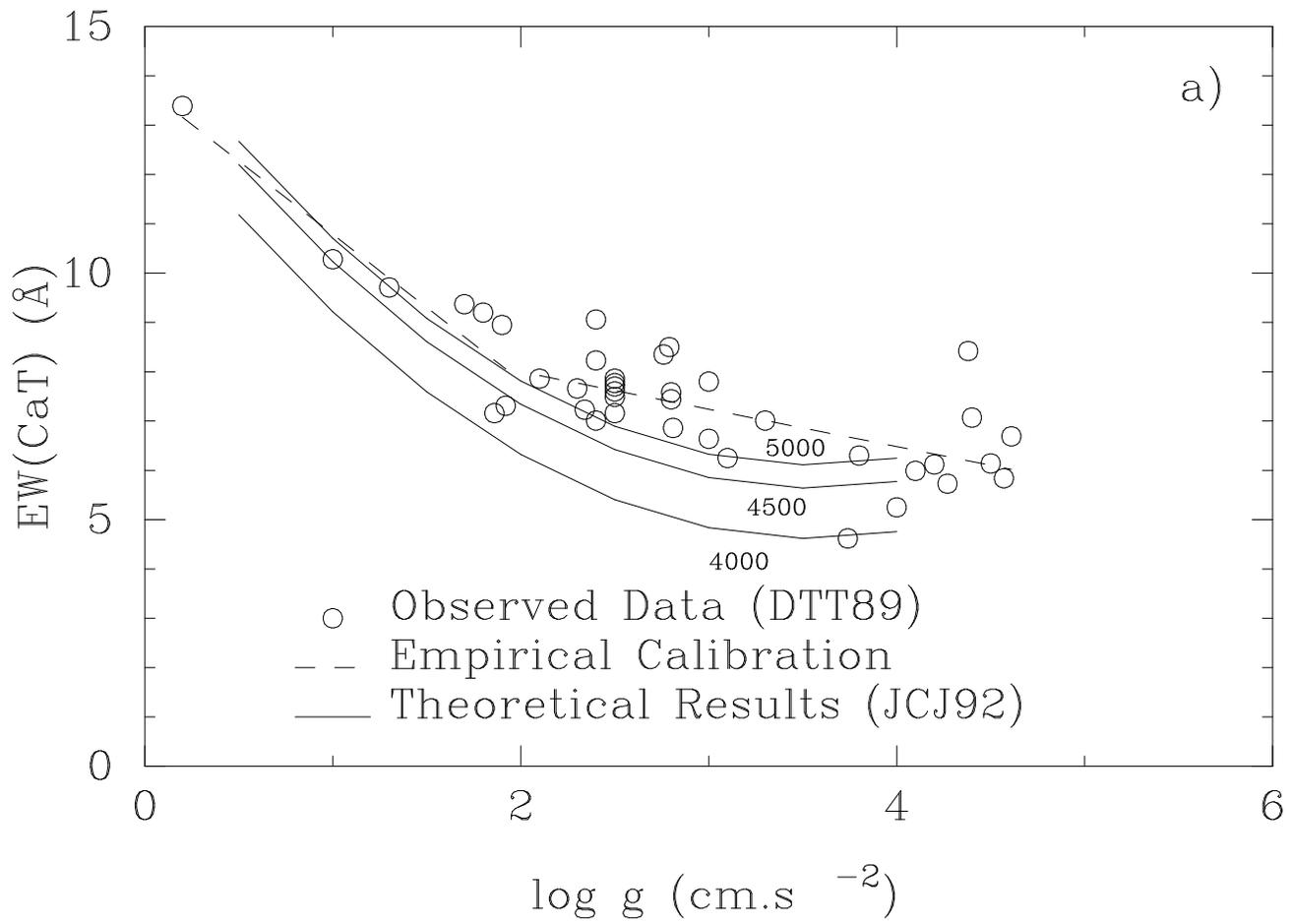
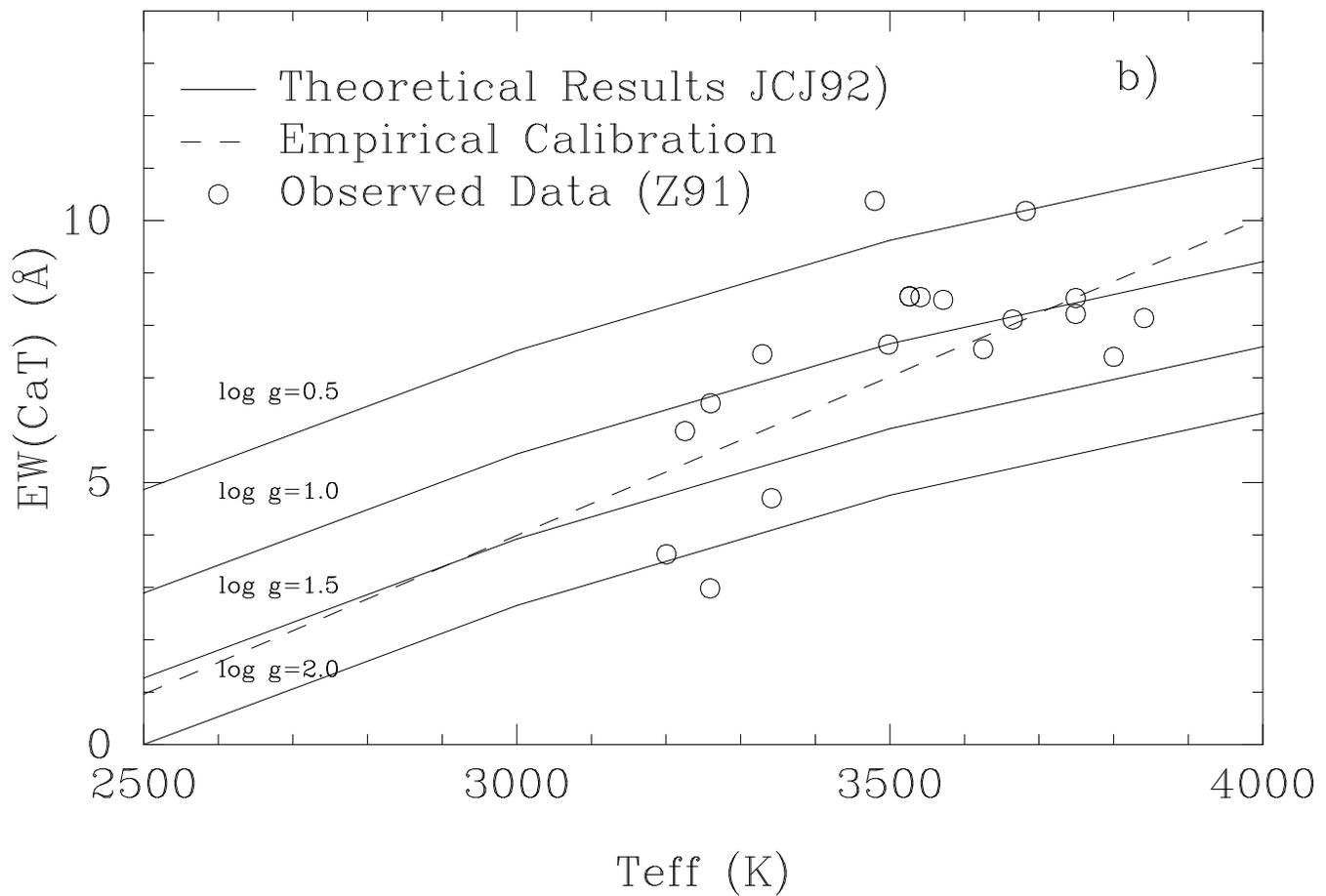
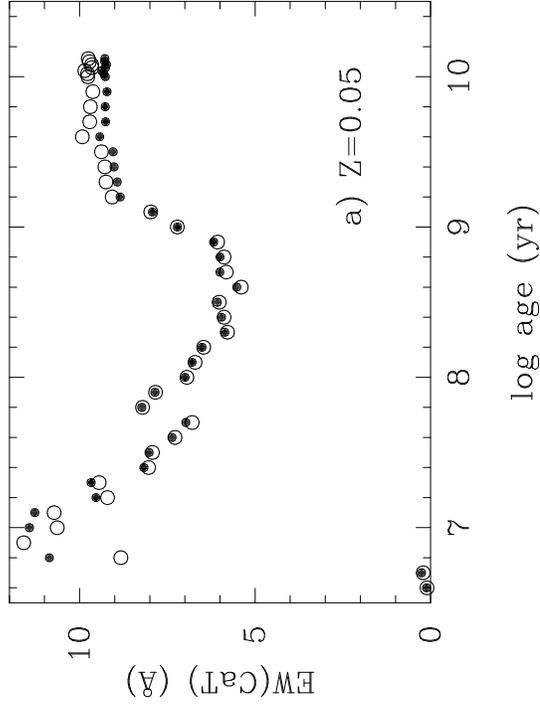
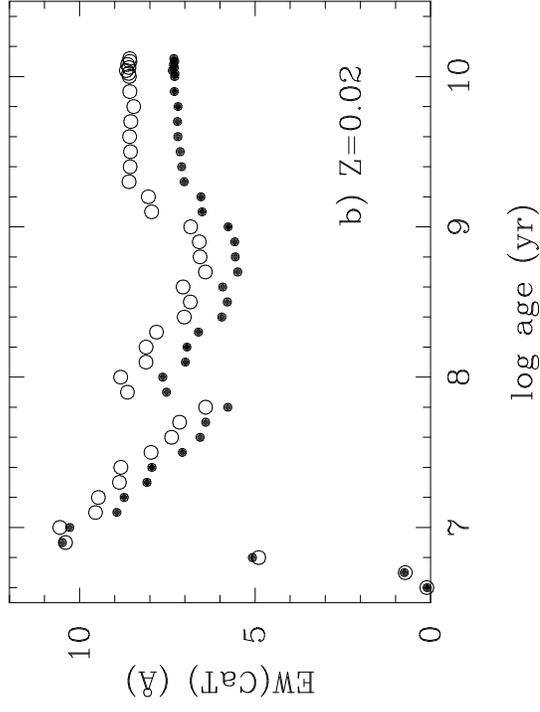
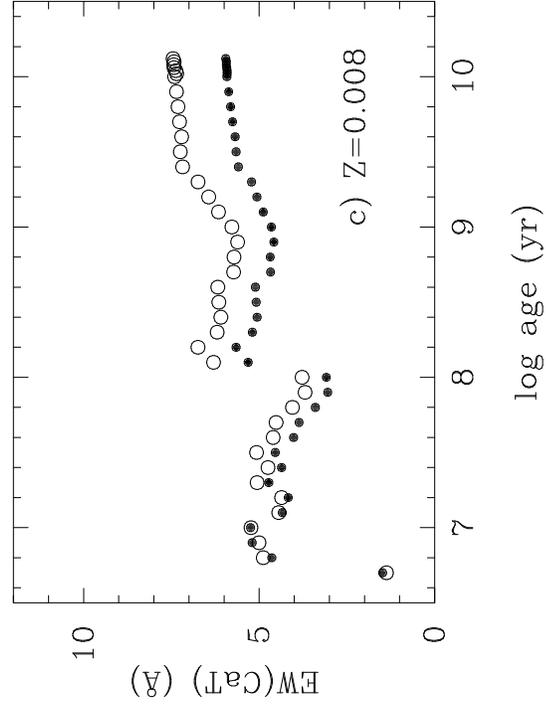
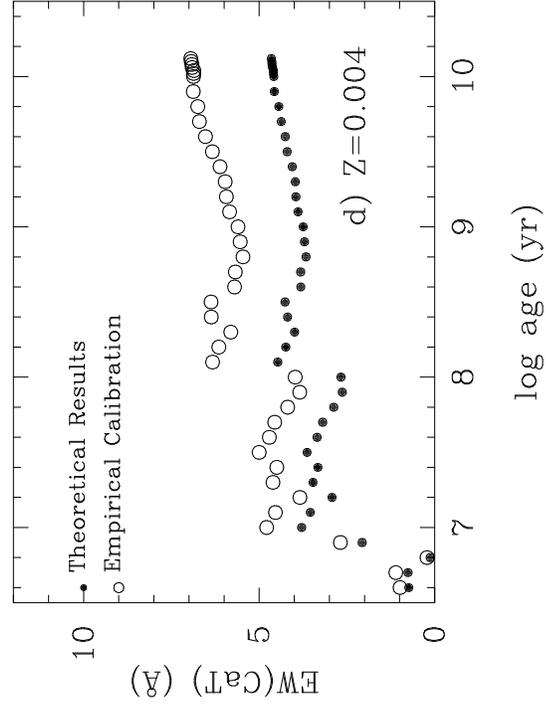
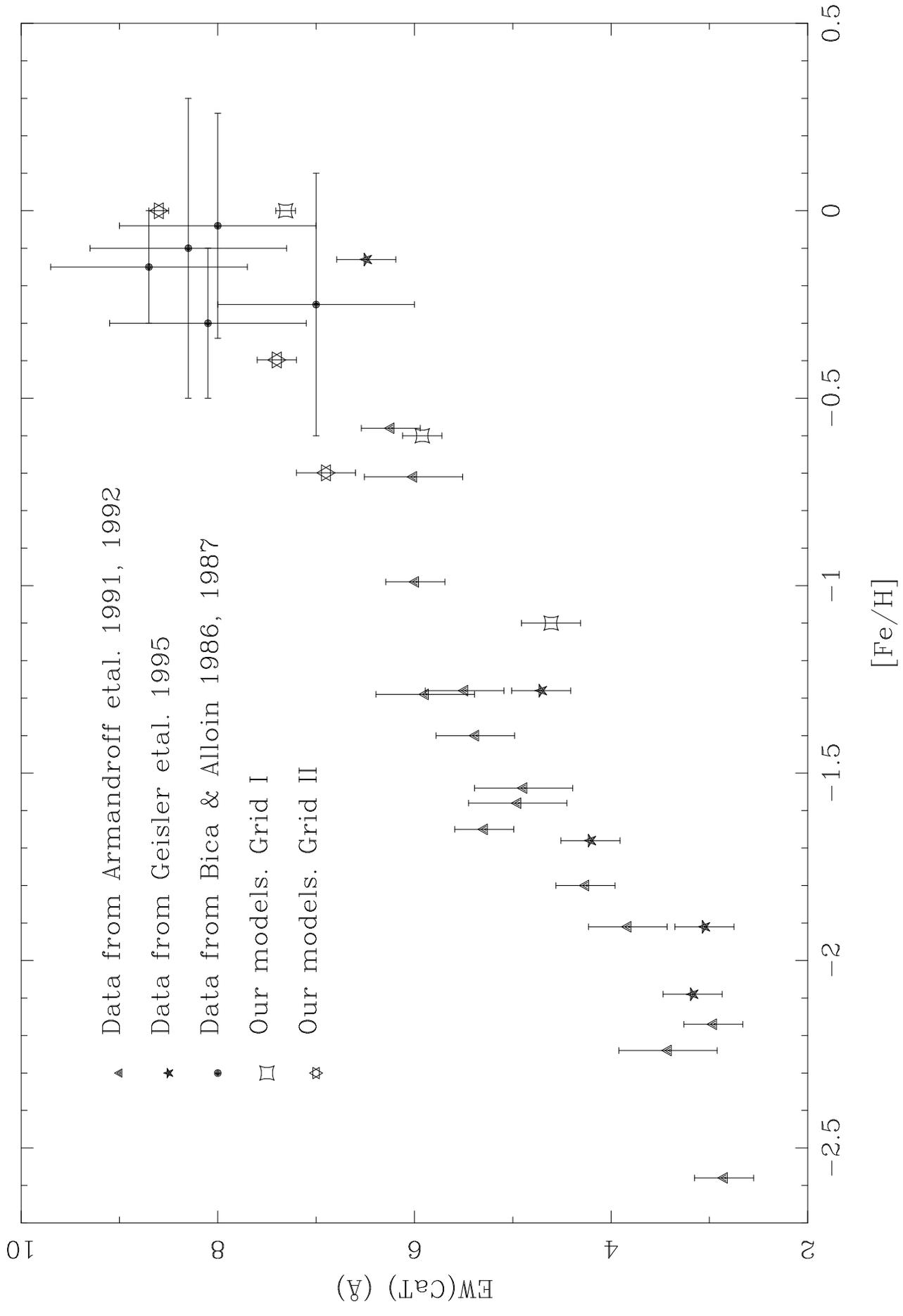
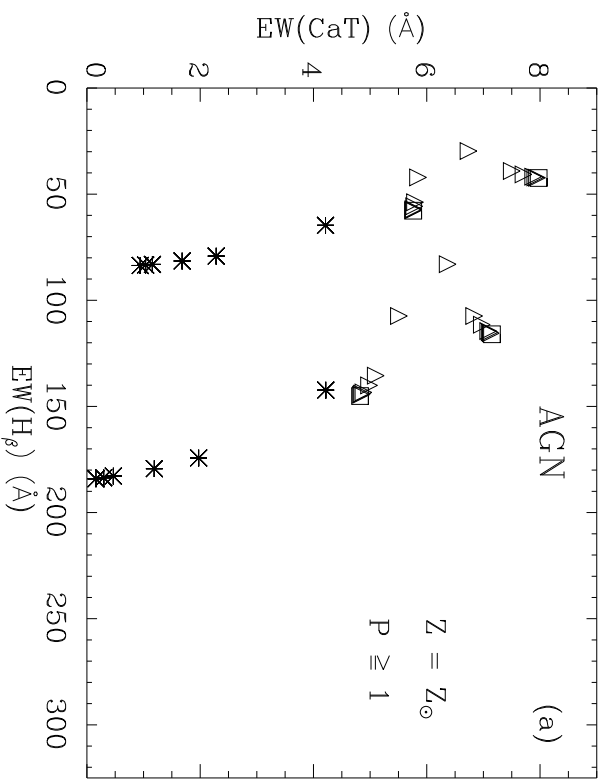
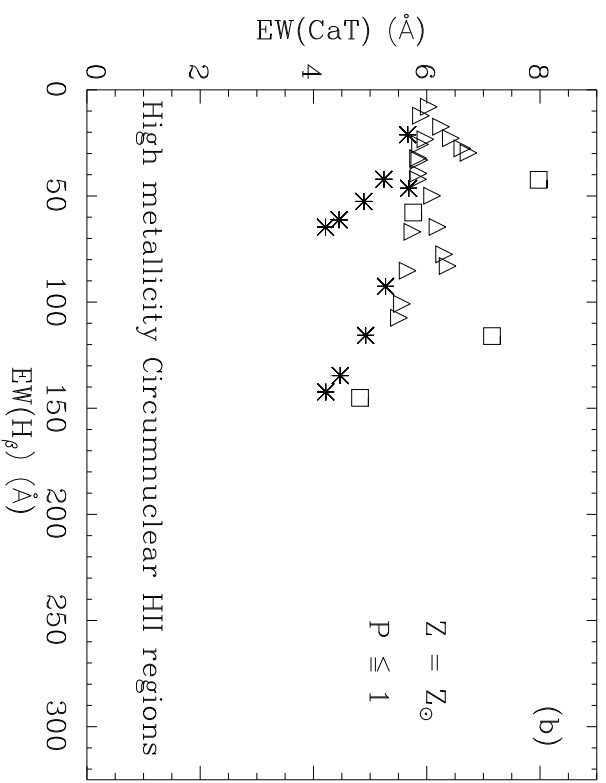
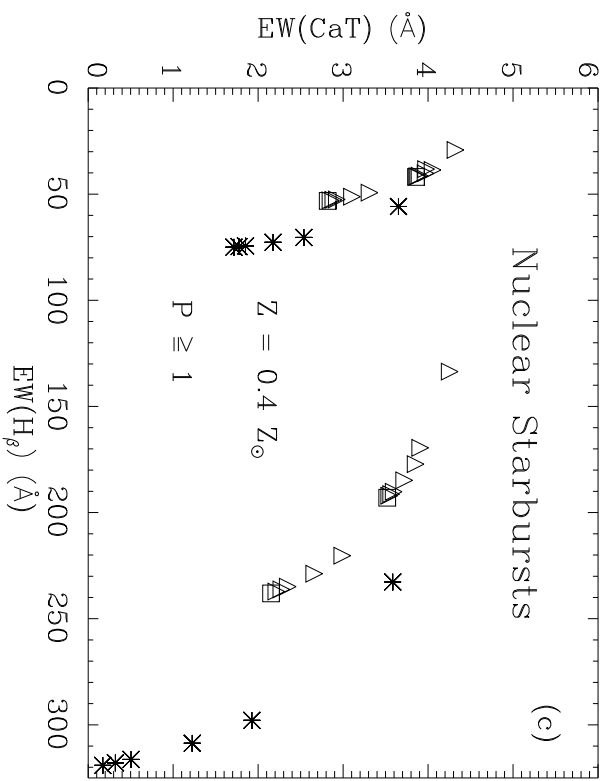
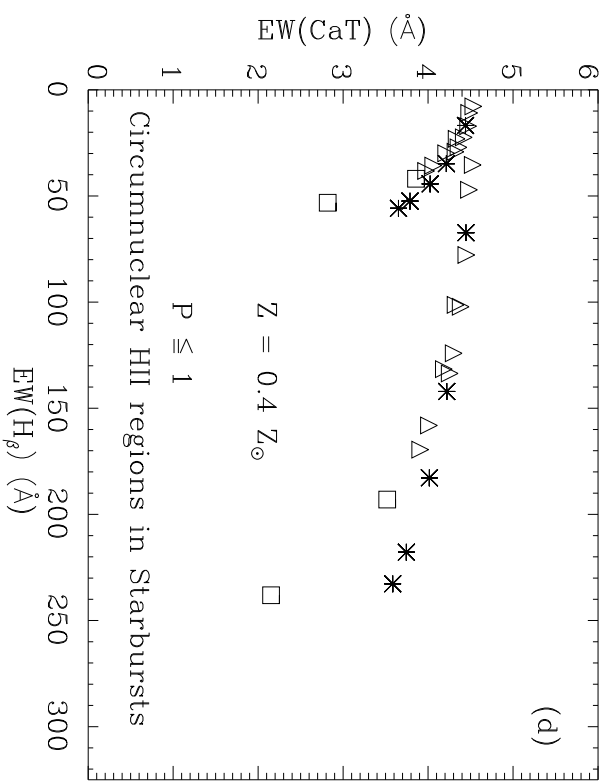
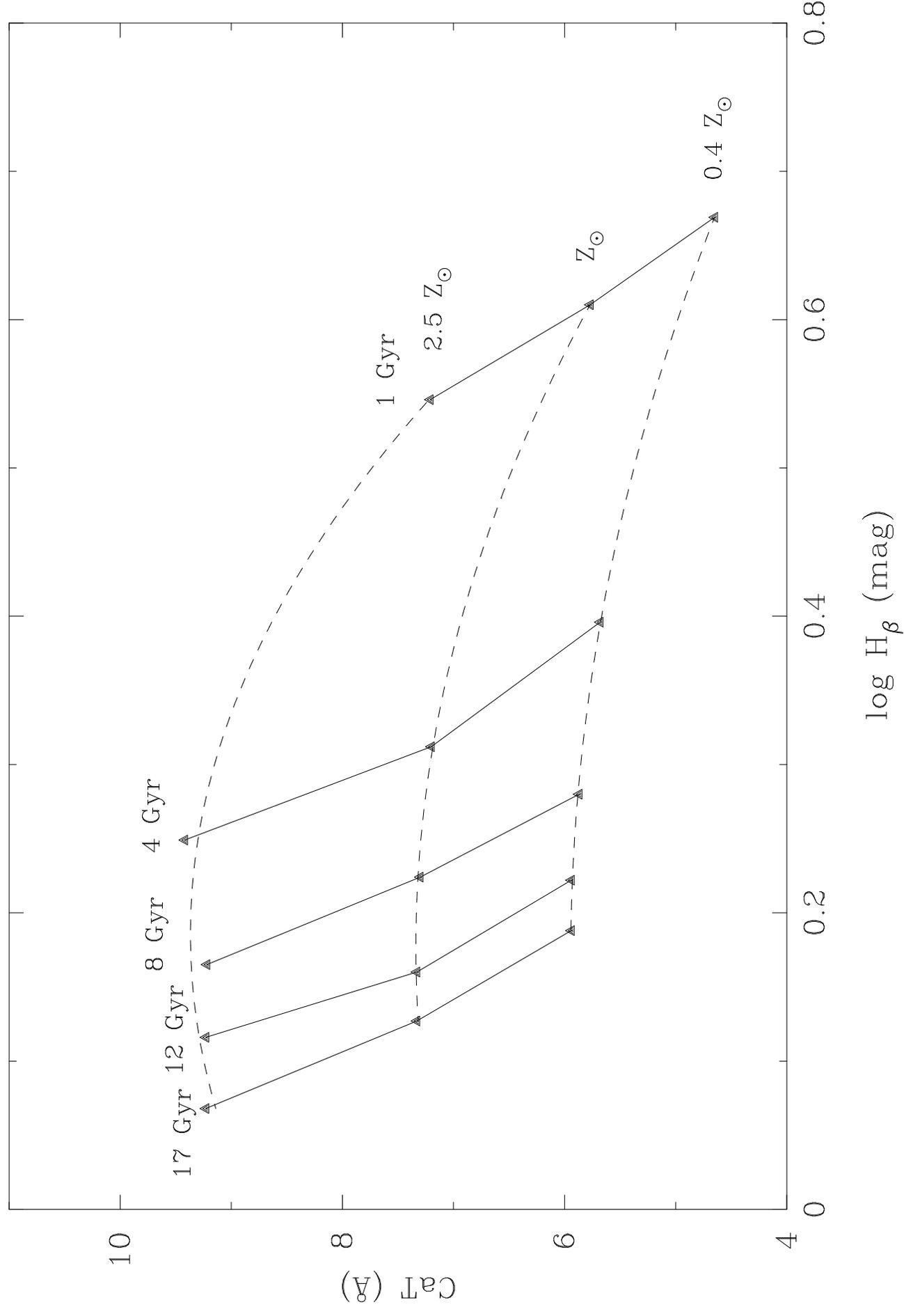
\end{document}